\documentclass[journal,twoside,twocolumn]{IEEEtran}
\pdfoutput=1  %

\usepackage{amsmath,amssymb}
\usepackage{graphicx}
\usepackage{xcolor}
\usepackage{cite}
\usepackage{balance}
\usepackage{hyperref}
\usepackage{subcaption} %
\usepackage{tikz}
\usepackage{tikz-3dplot}
\usetikzlibrary{arrows,positioning}

\hypersetup{
  breaklinks=true,  %
  colorlinks=true,  %
  pdfusetitle=true,  %
  citecolor=blue,  %
  linkcolor=blue  %
}

\newcommand\tstrut{\rule{0pt}{2.5ex}} %
\newcommand\bstrut{\rule[-1.3ex]{0pt}{0pt}} %

\newcommand{\E}{{\mathbb{E}}}
\newcommand{\R}{{\mathbb{R}}}
\newcommand{\Z}{{\mathbb{Z}}}
\newcommand{\T}{{\mathrm{T}}}
 
\newcommand{\ba}{{\boldsymbol{a}}}
\newcommand{\bA}{{\boldsymbol{A}}}
\newcommand{\bB}{{\boldsymbol{B}}}
\newcommand{\bD}{{\boldsymbol{D}}}
\newcommand{\bh}{{\boldsymbol{h}}}
\newcommand{\bH}{{\boldsymbol{H}}}
\newcommand{\bI}{{\boldsymbol{I}}}
\newcommand{\bM}{{\boldsymbol{M}}}
\newcommand{\bR}{{\boldsymbol{R}}}
\newcommand{\bu}{{\boldsymbol{u}}}
\newcommand{\bU}{{\boldsymbol{U}}}
\newcommand{\bv}{{\boldsymbol{v}}}
\newcommand{\bx}{{\boldsymbol{x}}}
\newcommand{\bX}{{\boldsymbol{X}}}
\newcommand{\bz}{{\boldsymbol{z}}}
\newcommand{\bzero}{{\boldsymbol{0}}}
\newcommand{\blambda}{{\boldsymbol{\lambda}}}
\newcommand{\bxi}{{\boldsymbol{\xi}}}

\newcommand{\baopt}{{\boldsymbol{a}_\mathrm{opt}}}

\newcommand{\Qopt}{{\hat{Q}}}
\newcommand{\Qsub}{{\tilde{Q}}}
\newcommand\Lamsub{{\tilde{\Lambda}}}
\newcommand{\bBp}{{\bB_\mathrm{p}}}
\newcommand{\Lamp}{{\Lambda_\mathrm{p}}}
\newcommand{\dx}{\,\mathrm{d}^n\bx}
\newcommand{\dxone}{\,\mathrm{d}^{n_1}\bx_1}

\newcommand{\dxi}{\,\mathrm{d}^{n_i}\bx_i}

\newcommand{\dxk}{\,\mathrm{d}^{n_k}\bx_k}

\DeclareMathOperator*{\argmin}{arg\,min} %
\DeclareMathOperator{\tr}{tr} %

\newcommand{\eq}[1]{\begin{align} #1 \end{align}}

\newtheorem{theorem}{Theorem}
\newtheorem{corollary}[theorem]{Corollary}
\newtheorem{prop}[theorem]{Proposition}
\newtheorem{lemma}[theorem]{Lemma}
\newcommand{\overlay}[3]{\makebox[0mm][l]{\hspace*{#1}\raisebox{#2}[0ex][0ex]{#3}}}

\title{On the Best Lattice Quantizers}
\author{Erik Agrell, \IEEEmembership{Fellow, IEEE}, and Bruce Allen, \IEEEmembership{Member, IEEE}
\thanks{Manuscript received 18 February 2022; revised 25 April 2023; accepted 17 June 2023.
The work of E.~Agrell was supported by a Collaborating Scientist Grant from the Max Planck Institute for Gravitational Physics, Germany, which is gratefully acknowledged.}%
\thanks{E.~Agrell is with the Department of Electrical Engineering, Chalmers University of Technology, 41296 Gothenburg, Sweden (e-mail: agrell@chalmers.se).}
\thanks{B.~Allen is with the Max Planck Institute for Gravitational Physics, 30167 Hannover, Germany, and
  Leibniz Universit\"at Hannover (e-mail: bruce.allen@aei.mpg.de).}
}

\begin{document}

\maketitle

\begin{abstract} %
  A lattice quantizer approximates an arbitrary real-valued source
  vector with a vector taken from a specific discrete lattice.
  The quantization error is the difference between the source vector
  and the lattice vector. In a classic 1996 paper, Zamir and Feder
  show that the globally optimal lattice quantizer (which minimizes the
  mean square error) has white quantization error: for a uniformly
  distributed source, the covariance of the error is the identity
  matrix, multiplied by a positive real factor.  We generalize the
  theorem, showing that the same property holds (i) for any lattice whose
  mean square error
  cannot be decreased by a small perturbation of the generator matrix,
  and (ii) for an optimal product of lattices that are themselves locally optimal in the sense of (i).
  We derive
  an upper bound on the normalized second moment (NSM) of the optimal
  lattice in any dimension, by proving that any lower- or
  upper-triangular modification to the generator matrix of a product
  lattice reduces the NSM.  Using these tools and employing the best
  currently known lattice quantizers to build product lattices, we
  construct improved lattice quantizers in dimensions 13 to 15, 17 to
  23, and 25 to 48. In some dimensions, these are the first reported
  lattices with normalized second moments below the best known upper bound.
\end{abstract}


\section{Introduction}

\IEEEPARstart{L}{attices} are regular arrays of points in $\R^n$. They
are obtained as arbitrary linear combinations of (at most $n$)
linearly independent basis vectors, with integer coefficients.  Hence,
lattices are a countably infinite set of vectors, closed under
addition. The remarkable book by Conway and Sloane \cite{splag}
provides a comprehensive review of lattices and their properties.

As fundamental geometric structures, lattices have found applications
in a variety of disciplines, including digital communications
\cite{zamir14book}, experimental design \cite{hamprecht03}, data
analysis \cite{PhysRevD.104.042005}, and particle physics
\cite{lerche89}. In each application, the problem of
designing the best lattice for a given purpose arises. Such optimization
challenges often reduce to familiar mathematical problems such as
sphere-packing, sphere-covering, or quantization
\cite[Ch.~1--2]{splag}.

In this paper, we are concerned with the \emph{quantization problem},
which can be defined as follows. Random vectors in $\R^n$ are drawn
from some (source) probability distribution, and approximated by their
closest lattice points. This approximation (or quantization) process
creates a round-off (or quantization) error: the difference between
the vector and its closest lattice point.  Among all lattices having
the same number of lattice points per unit volume, the optimal lattice
quantizer is the lattice with the minimum mean square error.  This is
equivalent to minimizing the \emph{normalized second moment} (NSM),
which is a scale-invariant measure of this mean square error.

As in most work on lattice quantization, we assume that the
lattice is sufficiently dense so that the source probability distribution
is approximately constant over each Voronoi region.
In this case, the optimal lattice does not depend upon the
source distribution of the random vectors.

Tables of the NSM, showing the best known lattices for quantization in
various dimensions are listed in \cite{gersho79, conway82voronoi},
\cite[p.~61]{splag}, and the quantization performance of some
additional lattices is computed in \cite{conway84, worley87, worley88,
  agrell98, dutour09, allen21}. Yet, proofs of optimality are known
only in dimensions up to three \cite{gersho79, barnes83}.

In a pioneering 1996 paper \cite{zamir96}, Zamir and Feder show that
the optimal lattice quantizer in any dimension has a \emph{white}
quantization error.  More precisely, the error defined above (vector
difference between a random source vector and its closest lattice
vector) has a covariance matrix which is the identity matrix, scaled
by a positive real constant.

In this paper, we extend the Zamir and Feder result to \emph{locally
optimal} lattices.  These are lattices whose NSM cannot be reduced by
a small perturbation of the lattice generator matrix.  We also
consider \emph{product lattices}, which are the Cartesian product of
two or more lower-dimensional lattices. The NSM of a product lattice
depends on the relative scaling between the component lattices.  
A closed-form expression for the optimal scale factors is derived,
and we call a product lattice using such scale factors an \emph{optimal product}.
If each of the lower-dimensional lattices is locally optimal, then we
prove that the optimal product is the
one for which the quantization error is white.

Lastly, we apply these methods to explicitly design some product
lattices and analytically optimize their scale factors. These provide
constructive upper bounds on the quantization performance of the
optimal lattices in their respective dimensions. This simple
construction yields better lattice quantizers than previously reported
in all dimensions above $12$ except for $16$ and $24$.  We also prove
that further optimization is possible: the NSM of such product lattices
is a saddle point in the space of generator matrices, and can be
further reduced by certain perturbations of the generator
matrix.

\section{Mathematical Preliminaries and Method} \label{s:prelim}

\emph{Notation:} Bold lowercase letters $\bx$ denote row vectors,
while bold uppercase letters $\bX$ denote either matrices or random
vectors. An all-zero vector or matrix of an arbitrary size (inferred
from the context) is denoted by $\bzero$, and identity matrices are
denoted by $\bI$. Sets are denoted by uppercase Greek letters $\Omega$, apart
from the integers $\Z$ and real numbers $\R$. Arithmetical operations
on sets should be understood as operating per element, e.g.,
$\Omega+\blambda \triangleq \{\bx+\blambda: \bx \in \Omega$\}.
Definitions are indicated by $\triangleq$.

Without loss of generality, we consider $n$-dimensional lattices
$\Lambda$ that are generated by square invertible $n \times n$
\emph{generator matrices} $\bB$.  The lattice consists of the set of
points $\bu \bB$ for all row vectors $\bu$ with integer components.
The all-zero row vector $\bzero$ belongs to all lattices.  The cubic
lattice $\Z^n$ is the special case for which $\bB$ is the identity
matrix.

Until now, we have used ``quantization'' to denote the map from a
vector in $\R^n$ to the closest lattice point.  However, for the proofs
in this paper, we consider more general mappings.  A
\emph{quantization rule} or \emph{quantizer} for a lattice $\Lambda$
is a function $Q_\Lambda(\bx)$ such that for all $\bx\in\R^n$
\eq{
Q_\Lambda(\bx) &\in \Lambda \label{e:cond1} , \\
Q_\Lambda(\bx+\blambda) &= Q_\Lambda(\bx)+\blambda
  \text{ for all } \blambda\in\Lambda \label{e:cond2} .
  }

The quantizer's properties are completely determined by
its behavior in the \emph{fundamental decision region} \eq{
  \Omega(Q_\Lambda) \triangleq \{\bx \in \R^n: Q_\Lambda(\bx) =
  \bzero\} \label{e:vqdef}, } since \eqref{e:cond2} may then be used to
determine the action anywhere.  The translate
$\Omega(Q_\Lambda)+\blambda $ of the fundamental decision region is
called the decision region of the lattice point $\blambda$.  All have the
same volume \cite[Prop.~2.2.1]{zamir14book}
\eq{ V_\Lambda \triangleq \int_{
    \Omega(Q_\Lambda)} \!\!\! \dx = |\!\det\! \bB| \label{e:volume} .
}
As indicated by the notation $V_\Lambda$, this volume depends upon the
lattice $\Lambda$, but is independent of the quantization rule.  Taken
together, the decision regions of all lattice points cover $\R^n$
without overlap.

As mentioned in the Introduction, the performance of lattice quantizers does
not depend upon the source distribution if the lattice is sufficiently dense. To prove this, consider 
a source probability density function (pdf)
$p_\bX(\bx)$, normalized by $\int p_\bX(\bx) \dx = 1$.
The mean square quantization error of the quantization rule is
\eq{ \label{e:qerror}
\E[\| \bx - Q_\Lambda(\bx)\|^2] &= \int_{\R^n} \!\!\! p_\bX(\bx) \| \bx - Q_\Lambda(\bx)\|^2 \dx .
}
Since the translates $ \Omega(Q_\Lambda)+\blambda, \forall
\blambda\in\Lambda$ cover $\R^n$ without overlap,
the mean square error can be written as
\eq{
\E[\| \bx - Q_\Lambda(\bx)\|^2] = \int_{ \Omega(Q_\Lambda)} \! \sum_{\blambda\in\Lambda} p_\bX(\blambda+\bx) \| \bx \|^2 \dx \label{e:mse},
}
where we use \eqref{e:cond2} to write this as an integral over the
fundamental decision region and \eqref{e:cond1} to set $Q_\Lambda(\bx)=\bzero$ inside
that region.
If now $\Lambda$ is sufficiently dense, then
$\sum_{\blambda\in\Lambda} p_\bX(\blambda+\bxi)$ is approximately
constant, independent of $\bxi$.  Such a probability distribution can
be obtained by rescaling a smooth base pdf $\tilde{p}$ of compact
support, for example as $p_\bX(\bx) = \alpha^n\tilde{p}(\alpha \bx)$
in the limit as $\alpha \rightarrow 0$.  In such a limit,
$\sum_{\blambda\in\Lambda}
p_\bX(\blambda+\bxi) = 1/V_\Lambda$ for any $\bxi$,\footnote{This can
be proved by writing $\int \tilde{p}(\bx)\dx = 1$ as a Riemann sum
over a shifted and scaled lattice $\bx \in \alpha(\Lambda+\bxi)$.}
and the mean square error \eqref{e:mse} approaches \cite{zador82}
\eq{
E(Q_\Lambda) & \triangleq \lim_{\alpha\rightarrow 0} \E[\|\bx-Q_\Lambda(\bx)\|^2] \label{e:eqlim} \\
&= \frac{1}{V_\Lambda} \int_{ \Omega(Q_\Lambda)} \!\!\! \| \bx \|^2 \dx \label{e:eqdef}.
}
Hence, in what follows, the pdf of the source does not appear.

Important quantities that are closely related to the mean square error are
the NSM or quantizer constant $G(Q_\Lambda)$ and the \emph{correlation matrix} $\bR(Q_\Lambda)$, which are \cite{gersho79}, \cite[pp.~48, 71]{zamir14book}
\eq{
G(Q_\Lambda) & \triangleq \frac{E(Q_\Lambda)}{nV_\Lambda^{2/n}} \label{e:gqdef}  , \\
\bR(Q_\Lambda) & \triangleq \frac{1}{V_\Lambda} \int_{ \Omega(Q_\Lambda)} \!\!\! \bx^\T \bx \dx \label{e:rqdef} .
}
Note that the NSM $G(Q_\Lambda)$ is ``dimensionless'' in the sense that it is
invariant under uniform rescaling of the lattice.
From \eqref{e:eqdef} and \eqref{e:rqdef}, it follows that
\eq{
E(Q_\Lambda) = \tr\!{\bR(Q_\Lambda)} \label{e:trace},
}
so the trace of the correlation matrix gives the mean square error.

It follows immediately from the definition \eqref{e:rqdef}
that the correlation matrix
$\bR$ is real, symmetric, and positive definite.  If
the quantization error is not white, then $\bR$ provides ``preferred
directions'' in the space, for example corresponding to the eigenvector
with the largest or the smallest eigenvalue. 
In the case of white
quantization error, however, $\bR$ is proportional to the identity, and does
not generate preferred directions, since every vector is an
eigenvector with the same positive real eigenvalue.

For a given lattice $\Lambda$, the most common and important rule is
the \emph{minimum-distance quantization rule}, denoted by a hat:
\eq{
  \label{e:mindist}
  Q_\Lambda(\bx) = \Qopt_\Lambda(\bx) \triangleq
  \argmin_{\blambda\in\Lambda} \|\bx-\blambda\|^2.
}
For any vector $\bx$, it returns the closest vector in the lattice.
Ties can be broken by any criterion that respects condition \eqref{e:cond2}.
This quantization rule is special
because, for a given lattice $\Lambda$, it minimizes $E(Q_\Lambda)$
and $G(Q_\Lambda)$.  This follows immediately from
\eqref{e:eqlim}, because the expectation
$\E[\|\bx-Q_\Lambda(\bx)\|^2]$ is minimal if
$\|\bx-Q_\Lambda(\bx)\|^2$ is minimized for every $\bx$.  Hence, $
\Qopt_\Lambda$ is the \emph{optimal decision rule} for a given lattice.

For this rule, the fundamental
decision region \eqref{e:vqdef} is the \emph{Voronoi region}
\eq{ \Omega(\Qopt_\Lambda) & \triangleq \{\bx \in \R^n: \| \bx \|^2
  \le \| \bx - \blambda \|^2, \;\forall \blambda \in \Lambda
  \} \label{e:vdef} , }
which geometrically consists of all points in $\R^n$ whose closest lattice point is the origin.%
\footnote{More precisely, the interiors of \eqref{e:vqdef} and
\eqref{e:vdef} are equal under the rule \eqref{e:mindist}. With a
slight abuse of notation, we disregard their boundaries, which have
zero $n$-volume and do not contribute to any integral over a finite
integrand.}
An important property of the Voronoi region of any lattice is that it
is symmetric about $\bzero$: the center of gravity 
$\int_{\Omega(\Qopt_\Lambda)} \bx \dx = \bzero$.  Hence, the correlation matrix
$\bR(Q_\Lambda)$ is equal to the
\emph{covariance matrix} whenever $Q_\Lambda = \Qopt_\Lambda$ (but not
for arbitrary quantization rules $Q_\Lambda$).

Throughout this paper, the word ``optimal'' is used in several
senses. For a given lattice, the \emph{optimal decision rule} is the one
which minimizes the NSM, i.e., \eqref{e:mindist}.  Among all lattices
of given dimension, the \emph{optimal lattice} is the one with the smallest
NSM.  The \emph{optimal product} of given lattices is the one that minimizes
the NSM among all Cartesian products of those lattices, by varying
the relative scales between them.

Our main theorem-proving technique is, as in \cite[Sec.~4.3]{zamir14book},
to construct different decision
rules for a given lattice, exploiting the fact that their NSMs are
equal to or greater than the NSM of the optimal decision rule $\Qopt$.
For example, in Sec.~\ref{s:local}, if the quantization error of a
lattice $\Lambda$ is not white, then $\bR$ provides preferred
directions
(say, the eigenvectors with the largest eigenvalue)\footnote{
A similar argument leads to \cite[Eq.~5.1.1]{Wald}.
}.  With
these, we construct a family of lattices $\Lamsub$ with nonoptimal
decision rules, whose NSM is smaller than that of the original lattice
$\Lambda$.  Since the NSM of $\Lamsub$ with an optimal decision rule
cannot be larger, we have thus shown that the original lattice is not
optimal. A similar proof technique is applied in Sec.~\ref{s:upper}. Starting with a
product lattice, we generate a new non-product lattice, with a
nonoptimal decision rule, but whose NSM is equal to that of the
original starting lattice.  Hence, the optimal decision rule on the new
non-product lattice must yield a smaller NSM than that of the original
product.

\section{Locally Optimal Lattices} \label{s:local}

Our starting point is the following theorem, which states that the
globally optimal quantizer lattice has a white quantization error: a
covariance matrix proportional to the identity.
\begin{theorem}[Zamir--Feder \cite{zamir96}, {\cite[Sec.~4.3]{zamir14book}}] \label{t:zf}
For the optimal lattice $\Lambda$ in any dimension $n$,
\eq{
\bR(\Qopt_\Lambda) = \frac{E(\Qopt_\Lambda)}{n}\bI . \label{e:white}
}
\end{theorem}
We now generalize this to the locally optimal case.

A \emph{locally optimal lattice} is a lattice $\Lambda$ whose NSM
$G(\Qopt_\Lambda)$ cannot be
decreased by an infinitesimal perturbation of the generator matrix
$\bB$ \cite{barnes83}. The extension to Theorem \ref{t:zf} is:

\begin{theorem} \label{t:local}
Any locally optimal lattice satisfies \eqref{e:white}.
\end{theorem}

\begin{IEEEproof}
  Our proof is constructive. If the covariance matrix of $\Lambda$ is
  not proportional to the identity matrix, we use it to build a nearby
  lattice $\Lamsub$ with a smaller NSM than the NSM of $\Lambda$.

Let $\Lamsub = \Lambda \bA_\beta$, where $\bA_\beta$ is an invertible
$n \times n$ matrix and $\beta$ is a real parameter to be defined
later.  As in \cite{zamir96}, we consider the minimum-distance
quantizer $\Qopt_{\Lambda}(\bx)$ on $\Lambda$ and the suboptimal
quantizer $\Qsub_\Lamsub(\bx) \triangleq
\Qopt_\Lambda(\bx\bA_\beta^{-1}) \bA_\beta$ on $\Lamsub$.

It is straightforward to show that $\Qsub_\Lamsub(\bx)$ satisfies
\eqref{e:cond1}--\eqref{e:cond2} and has a fundamental decision
region $ \Omega(\Qsub_\Lamsub) = \Omega(\Qopt_\Lambda) \bA_\beta$.
Note that while $\Omega(\Qopt_\Lambda)$ is the Voronoi region of
$\Lambda$, the fundamental decision region $\Omega(\Qsub_\Lamsub)$ is
generally \emph{not} the Voronoi region of $\Lamsub$. By
\eqref{e:volume}, it has volume $V_\Lamsub =
V_\Lambda\,|\!\det\!\bA_\beta|$.

The covariance matrices of the two quantization rules are easily related
using the change of variables (mapping) provided by $\bA_\beta$.  From
\eqref{e:rqdef}, the covariance matrix of $\Qsub_\Lamsub$ is
$\bR(\Qsub_\Lamsub) = \bA_\beta^\T \bR(\Qopt_\Lambda) \bA_\beta$
\cite[Eq.~(15)]{zamir96}.
Hence, the NSM is 
\eq{ \label{e:gbeta}
G(\Qsub_\Lamsub) = \frac{E(\Qsub_\Lamsub)}{n V_\Lamsub^{2/n}} 
= \frac{\tr (\! \bA_\beta \!  \bA_\beta^T \! \bR(\Qopt_\Lambda))}{n(V_\Lambda\,|\!\det\!\bA_\beta|)^{2/n}}, 
}
where we have used the cyclic property of the trace.

To select $\bA_{\beta}$, we follow the approach described earlier, using
$\bR(\Qopt_\Lambda)$
to obtain preferred directions.%
\footnote{From here on, our proof deviates from the corresponding proof in \cite{zamir96} for \emph{globally} optimal lattices.}
Let $\bar \bR$ denote the traceless part, which by assumption is nonzero:
\eq{ \label{e:rbar}
  \bar \bR \triangleq \bR(\Qopt_\Lambda) - \frac{\tr \! \bR(\Qopt_\Lambda)}{n} \bI ,
}
and let $\bA_\beta \triangleq \exp(\beta \bar \bR)$.
This choice of mapping is volume-preserving, since for any square matrix
$\bM$, $\det(\exp(\bM)) = \exp( \tr \bM)$ \cite[p.~16]{perko01}.  Thus, $\det\!\bA_\beta =
\exp(\beta \tr \bar \bR) = 1$.
Note that because the covariance matrix is symmetric and real, both $\bar \bR$ and
$\bA_\beta$ are symmetric and real.

For the proof, we only need $\bA_\beta$ for infinitesimal $\beta$:
\eq{\bA_\beta = \bI + \beta \bar \bR + O(\beta^2). \label{e:abeta}}
Substituting $\bA_\beta$ from \eqref{e:abeta} and
$\bR(\Qopt_\Lambda) =  \bI \tr \! \bR(\Qopt_\Lambda)/n + \bar \bR$ 
from \eqref{e:rbar}
into \eqref{e:gbeta}, the NSM becomes
\begin{align}  \label{e:gbeta2}
  G(\Qsub_\Lamsub) & = \frac{1}{n V_\Lambda ^{2/n}} \tr\! \left(\left[\bI + \beta \bar \bR + O(\beta^2)\right]^2  \left[\frac{\tr\!\bR(\Qopt_\Lambda)}{n} \bI + \bar{\bR} \right] \right) \notag\\
            &= G(\Qopt_\Lambda) + 2\beta \frac{\tr \bar{\bR}^2 }{n V_\Lambda^{2/n}} + O(\beta^2),
\end{align}
where we have distributed the trace over additions and used $\tr\!\bar\bR = 0$.

It is clear from \eqref{e:gbeta2} that for negative $\beta$ near zero, we have
$G(\Qsub_\Lamsub) < G(\Qopt_\Lambda)$.  This follows because, since
$\bar \bR$ is a non-vanishing real symmetric matrix, $ \tr \bar \bR^2$
must be positive\footnote{To prove this, write $\bar \bR = \bU \bD
\bU^{-1}$ where $\bU$ is orthogonal and $\bD$ is real and diagonal,
then use the cyclic property of the trace.}.
Since the NSM of the minimum-distance
quantization rule \eqref{e:mindist} on $\Lamsub$ satisfies\footnote{
We have no way to directly analyze the performance
of $\Qopt_\Lamsub$ because we have no simple expression for the
Voronoi region $ \Omega(\Qopt_\Lamsub)$.}
$G(\Qopt_\Lamsub) \le G(\Qsub_\Lamsub)$, we
have established that for negative $\beta$ near zero,
$G(\Qopt_\Lamsub) < G(\Qopt_\Lambda)$.
\end{IEEEproof}

To test Theorem \ref{t:local}, we examine a large number of
numerically optimized lattice quantizers. These were designed in 1996
using an iterative algorithm, which converges to different locally
optimal lattices \cite{agrell98}. A total of $90$ locally optimal
lattices are available as online supplementary material to the 1998
article \cite{agrell-supplements}; we
estimate the covariance matrices $\bR(\Qopt_\Lambda)$ of their
quantization errors using Monte Carlo integration. In all cases,
consistent with the theorem, the obtained covariance matrices are
proportional to the identity matrix, apart from minor round-off
errors.

Theorem \ref{t:local} establishes that local optimality is a sufficient
condition for a white quantization error \eqref{e:white}.
Is it also a necessary condition? In other words,
is any lattice that satisfies \eqref{e:white} locally optimal?
In the next two sections, we will show that this is false, using
product lattices as a counterexample.

\section{Product Lattices} \label{s:product}

In this section, we study lattices that are formed as the Cartesian
products of lower-dimensional lattices.
Gersho applied this technique to obtain
upper bounds on the optimal NSM for $n=5$ and $n=100$, without
formalizing the expressions \cite[Sec.~VII]{gersho79}.

Let $k$ lattices in dimensions $n_1,\ldots,n_k$ be denoted by
$\Lambda_1,\ldots,\Lambda_k$, and consider their product $\Lamp =
\Lambda_1 \times \cdots \times \Lambda_k$, whose dimension is $n = n_1+\cdots+n_k$. A
generator matrix for $\Lamp$ is
\eq{ \label{e:bp} \bBp
  = \begin{bmatrix}
    \bB_1 & \cdots & \bzero \\
    \vdots & \ddots & \vdots \\
    \bzero & \cdots & \bB_k
\end{bmatrix} ,
}
where $\bB_i$ is a generator matrix of $\Lambda_i$ for $i=1,\ldots,k$.
The Voronoi region, volume, and other properties of a product lattice
are as follows.
\begin{prop} \label{t:prod}
For any $k \ge 1$ and $\Lamp = \Lambda_1 \times \cdots \times \Lambda_k$,
\eq{
 \Omega & \triangleq
  \Omega(\Qopt_{\Lamp})=  \Omega_1 \times \cdots \times \Omega_k \label{e:vprod} , \\
V &\triangleq V_{\Lamp} = V_1 V_2 \cdots V_k \label{e:volprod} , \\
E &\triangleq E(\Qopt_{\Lamp}) = E_1+\cdots+E_k \label{e:eprod} , \\
G &\triangleq G(\Qopt_{\Lamp}) = \frac{1}{n V^{2/n}} \sum_{i=1}^k n_i V_i^{2/n_i} G_i \label{e:gprod} , \\
\bR &\triangleq
  \bR(\Qopt_{\Lamp}) = \begin{bmatrix}
  \bR_1 & \cdots & \bzero \\
  \vdots & \ddots & \vdots \\
  \bzero & \cdots & \bR_k
\end{bmatrix} \label{e:rprod} ,
}
where $\Omega_i$, $V_i$, $E_i$, $G_i$, and $\bR_i$ denote the
corresponding properties of $\Lambda_i$.
\end{prop}

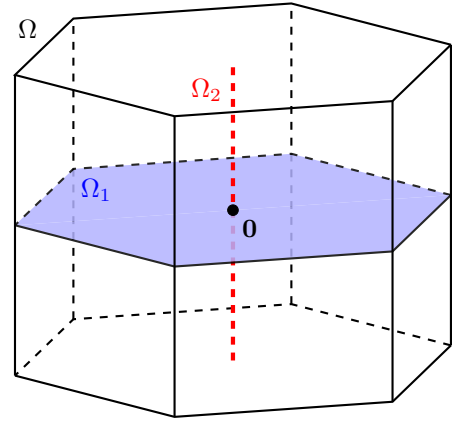
\begin{figure}
\begin{center}
\tdplotsetmaincoords{75}{-15} %
\begin{tikzpicture}[tdplot_main_coords]
\def\radius{3}
\def\height{20mm}

\begin{scope}[yshift=-\height]
\draw[thick] (\radius,0)
  \foreach \x in {0,300,240,180} { --  (\x:\radius) node (N1-\x) {} };
\draw[dashed,thick] (N1-0.center)
  \foreach \x in {60,120,180} { --  (\x:\radius) node (N1-\x) {} };
\node[at=(0:0)] (F1) {};
\end{scope}

\begin{scope}[yshift=\height]
\draw[thick] (\radius,0)
  \foreach \x in {0,60,120,180,240,300,360} { -- (\x:\radius) node (N2-\x) {}};
\node[at=(0:0)] (F2) {};
\node[at=(135:1.05*\radius)] {$ \Omega$};
\end{scope}

\foreach \x in {60,120} { \draw[thick,dashed] (N1-\x.center)--(N2-\x.center); };
\draw[thick,dashed,fill=blue!30,opacity=.85] (0:\radius)
  \foreach \x in {60,120,180} { -- (\x:\radius) };
\node[blue,at=(135:.7*\radius)] {$ \Omega_1$};

\draw[dashed,ultra thick,red] (F1.center) -- (F2.center) node[pos=.9, left] {$ \Omega_2$};

\draw[thick,fill=blue!30,opacity=.85] (0:\radius)
  \foreach \x in {300,240,180} { -- (\x:\radius) };
\foreach \x in {0,180,240,300} { \draw[thick] (N1-\x.center)--(N2-\x.center); };

\filldraw[black] (0:0) circle (2pt) node[anchor=north west]{$\bzero$};

\end{tikzpicture}
\end{center}
\caption{\label{f:topology} The Voronoi region $ \Omega$ of the
  product lattice $\Lamp = \Lambda_1 \times \Lambda_2$, where
  $\Lambda_1$ is the two-dimensional hexagonal lattice $A_2$ and
  $\Lambda_2$ is the one-dimensional integer lattice $\Z$.  The origin
  $\bzero$ belongs to all three lattices and is the centroid of all
  three Voronoi regions.  The top and bottom facets of $ \Omega$ are
  shifted copies of $ \Omega_1$, and the six vertical edges are
  shifted copies of $ \Omega_2$.}
\end{figure}

An example of a $3$-dimensional Voronoi region $ \Omega$, constructed
according to Proposition \ref{t:prod}
as the Cartesian product of two lower-dimensional Voronoi regions $
\Omega_1$ and $ \Omega_2$, is illustrated in Fig.~\ref{f:topology}.

\begin{IEEEproof}
  Let $\bx = [ \bx_1 \;\cdots\; \bx_k ]$ and $\blambda = [
    \blambda_1 \;\cdots\; \blambda_k ]$. By definition, $ \Omega$ in \eqref{e:vdef} is
  formed by all vectors $\bx_i \in \R^{n_i}$ for $i=1,\ldots,k$
  such that
\eq{ \label{e:vineq}
 \| \bx_1 \|^2 +\cdots+ \| \bx_k \|^2 &\le \| \bx_1 - \blambda_1 \|^2 +\cdots+ \| \bx_k - \blambda_k \|^2,
}
for all $\blambda_i \in \Lambda_i$.

If $\bx_i\in \Omega_i$ for all $i$, then $ \| \bx_i \|^2
\le \| \bx_i - \blambda_i \|^2$.
Summing over $i=1,\ldots,k$ implies \eqref{e:vineq}, so $\Omega_1
\times\cdots\times \Omega_k \subseteq \Omega$.
Conversely, if $\bx\in \Omega$, then for a fixed $i$ we set $\blambda_j = \bzero$ in
\eqref{e:vineq} for all $j\ne i$. This implies $ \| \bx_i \|^2 \le \| \bx_i - \blambda_i
\|^2$ for all $\blambda_i \in \Lambda_i$, so that $\bx_i \in
\Omega_i$.  Repeating for $i=1,\ldots,k$ shows that
$\Omega \subseteq \Omega_1 \times\cdots\times \Omega_k $.

Taken together, $\Omega_1 \times\cdots\times \Omega_k \subseteq \Omega$ and
$\Omega \subseteq \Omega_1 \times\cdots\times \Omega_k $ prove \eqref{e:vprod},
which in turn proves \eqref{e:volprod}.

The definition \eqref{e:eqdef}, applied to a product lattice using 
\eqref{e:vprod} and \eqref{e:volprod}, implies that
\eq{
E
\nonumber
&= \frac{1}{V} \int_{\Omega_1} \!\!\cdots\! \int_{\Omega_k} \!\!\! \left( \| \bx_1 \|^2 +\cdots+ \| \bx_k \|^2 \right) \dxone \cdots \dxk \\
&= \frac{1}{V_1 V_2\cdots V_k} \left( (V_1 E_1) V_2\cdots V_k +\cdots+ V_1 V_2 \cdots (V_k E_k)  \right), \label{e:eprod2}
}
which proves \eqref{e:eprod}.
Equation \eqref{e:gprod} follows by substituting $E = n V^{2/n} G$
and the corresponding expressions for $E_i$ into
\eqref{e:eprod}, and simplifying using \eqref{e:volprod} and $n=n_1 +\cdots+
n_k$.

Lastly, to prove \eqref{e:rprod}, we use \eqref{e:vprod} in \eqref{e:rqdef} to obtain
\eq{
\bR
&= \frac{1}{V} \int_{\Omega_1} \!\!\cdots\! \int_{\Omega_k} \!
\begin{bmatrix}
\bx_1^\T \bx_1 & \cdots & \bx_1^\T \bx_k \\
\vdots & \ddots & \vdots \\
\bx_k^\T \bx_1 & \cdots & \bx_k^\T \bx_k
\end{bmatrix}
\dxone \cdots \dxk . \label{e:rprod2}
}
For the submatrices on the diagonal, whose integrands have the form $\bx_i^\T \bx_i$, the volumes cancel as in \eqref{e:eprod2}, leaving $\bR_i$. The off-diagonal submatrices with integrands $\bx_i^\T \bx_j$ for $i \ne j$ are separable into products of two integrals such as $\int_{\Omega_i} \bx_i \dxi$.
These vanish because (as pointed out after \eqref{e:vdef}) the Voronoi
region $\Omega_i$ is symmetric about zero and thus has its
center of gravity at the origin.
\end{IEEEproof}

We now generalize the product construction by introducing a list of real positive scale factors
$\ba = [a_1,\ldots,a_k]$ to build a family of product
lattices $\Lambda(\ba) = a_1\Lambda_1 \times\cdots\times a_k\Lambda_k$. A generator
matrix for $\Lambda(\ba)$ is
\eq{ \bB(\ba) = \begin{bmatrix}
  a_1 \bB_1 & \cdots & \bzero \\
  \vdots & \ddots & \vdots \\
  \bzero & \cdots & a_k \bB_k
\end{bmatrix} \label{e:bproda} .
}
The properties of $\Lambda(\ba)$ follow by
replacing $\Omega_i$ by $a_i \Omega_i$, $V_i$ by
$a_i^{n_i} V_i$, $E_i$ by $a_i^2 E_i$, and $\bR_i$ by $a_i^2 \bR_i$
in Proposition \ref{t:prod}, while
$G_i$, due to its scale-invariant definition \eqref{e:gqdef}, remains unchanged
for $i=1,\ldots,k$.
These substitutions result in
\eq{
 \Omega(\ba) &=  a_1\Omega_1 \times\cdots\times a_k \Omega_k , \\
V(\ba) &= a_1^{n_1}\cdots a_k^{n_k} V_1 \cdots V_k , \label{e:vproda} \\
E(\ba) &= a_1^2 E_1+\cdots+a_k^2 E_k \label{e:eproda} , \\
G(\ba) &= \frac{1}{n V(\ba)^{2/n}} \sum_{i=1}^k n_i a_i^2 V_i^{2/n_i} G_i \label{e:gproda} , \\
\bR(\ba) &= \begin{bmatrix}
  a_1^2 \bR_1 & \cdots & \bzero \\
  \vdots & \ddots & \vdots \\
  \bzero & \cdots & a_k^2\bR_k
\end{bmatrix} \label{e:rproda} .
}

For given $\Lambda_1, \ldots, \Lambda_k$, what scale factors $\ba$
produce the \emph{optimal product} $\Lambda(\ba)$, in the sense
of minimizing $G(\ba)$?
We note that $G(\ba)$ has at least one minimum
for finite and positive $a_1,\ldots,a_k$, because if one of these scale factors
is varied while keeping the others fixed, then \eqref{e:gproda} diverges
to infinity as the scale factor tends to either zero or infinity.
This minimum is unique up to a linear scale factor, and has a closed
form as follows.
\begin{theorem} \label{t:optprod}
For given lattices $\Lambda_1,\ldots,\Lambda_k$, varying only $\ba$,
$\Lambda(\ba)$ is an optimal product if and only if
\eq{
a_i = \frac{C}{V_i^{1/n_i}\sqrt{G_i}}, \quad i = 1,\ldots, k \label{e:aoptgg}
}
for an arbitrary real constant $C>0$. An equivalent condition is
\eq{
a_i = C \sqrt{\frac{n_i}{E_i}}, \quad i = 1,\ldots, k \label{e:aoptee} .
}
The optimal NSM $G(\ba)$ is given by
\eq{
G^n(\ba) = G_1^{n_1} \cdots G_k^{n_k} \label{e:goptgg}
}
independently of $V_1,\ldots, V_k$.
\end{theorem}

\begin{IEEEproof}
The derivative of $G(\ba)$ in \eqref{e:gproda} with respect to $a_i$ is
\eq{
\frac{\partial G(\ba)}{\partial a_i} &= \frac{1}{n V(\ba)^{2/n}} \cdot 2n_i a_i V_i^{2/n_i} G_i \notag\\
&\quad - \frac{2}{n^2 V(\ba)^{1+2/n}} \frac{\partial V(\ba)}{\partial a_i} \cdot \sum_{j=1}^k n_j a_j^2 V_j^{2/n_j} G_j \label{e:dga1} .
}
Substituting
\eq{
\frac{\partial V(\ba)}{\partial a_i} = \frac{n_i}{a_i}V(\ba) ,
}
which follows from \eqref{e:vproda}, into \eqref{e:dga1} and simplifying yields
\eq{
\frac{\partial G(\ba)}{\partial a_i} = \frac{2n_i}{n V(\ba)^{2/n}} \left( a_i V_i^{2/n_i} G_i - \frac{1}{n a_i} \sum_{j=1}^k n_j a_j^2 V_j^{2/n_j} G_j \right) \label{e:dga2}.
}
Equating \eqref{e:dga2} to zero for $i=1,\ldots,k$ reveals that $a_i^2 V_i^{2/n_i} G_i$
is constant for all $i$, which gives \eqref{e:aoptgg}.
Substituting \eqref{e:aoptgg} back into \eqref{e:dga2} confirms that
\eqref{e:aoptgg} is not only necessary for \eqref{e:dga2} being zero but also sufficient,
regardless of $C$.
Then \eqref{e:aoptee} follows from \eqref{e:aoptgg}
and \eqref{e:gqdef}.
Substitute \eqref{e:aoptgg} into \eqref{e:vproda} and \eqref{e:gproda},
and use $n_1+\cdots+n_k = n$. Finally, $V_i$ and $C$ cancel out and \eqref{e:goptgg} emerges.
\end{IEEEproof}

An interesting special case is when the sublattices $\Lambda_i$
are locally optimal for all $i=1,\ldots,k$.
If the scale factors $\ba$ are optimally chosen according to Theorem \ref{t:optprod},
which we denote by $\baopt$, then $\Lambda(\baopt)$ also has white quantization error.
We state this in a way similar to
Theorems \ref{t:zf} and \ref{t:local}, but with different conditions.
\begin{corollary} \label{t:ropt}
If $\Lambda_1,\ldots,\Lambda_k$ are locally optimal lattices,
$\Lambda(\ba) = a_1 \Lambda_1 \times \cdots \times a_k \Lambda_k$,
and $[a_1,\ldots,a_k] = \baopt$, then
\eq{
\bR(\baopt) = \frac{E(\baopt)}{n}\bI \label{e:ropt} .
}
Furthermore, $G(\baopt)$ is locally minimal with respect to any perturbations in
the submatrices on the diagonal of \eqref{e:bproda}.
\end{corollary}
\begin{IEEEproof}
From Theorem \ref{t:local}, $\bR_i = (E_i/n_i)\bI$ for $i = 1,\ldots,k$.
Using this together with \eqref{e:aoptee} in
\eqref{e:rproda} yields
$\bR(\baopt) = C^2 \bI = (E(\baopt)/n) \bI$, where the latter equality is proved by substituting
\eqref{e:aoptee} into \eqref{e:eproda}.

To prove the local optimality of $G(\baopt)$, we let $a'_i \triangleq a_i V_i^{1/n_i}$
and $\bB'_i \triangleq \bB_i V_i^{-1/n_i}$, where as before $V_i = \det\! \bB_i$. Then the submatrices on the diagonal of \eqref{e:bproda} can be written as $a_i \bB_i = a'_i \bB'_i$ for all $i$, where $V'_i \triangleq \det\! \bB'_i = 1$. We will consider variations in $a'_i$ and $\bB'_i$ separately.
First, if $a'_i$ is varied for any fixed (not necessarily optimal) $\bB'_i$, then Theorem \ref{t:optprod} applies and the minimal NSM is attained when $a'_i = C/\sqrt{G_i}$.
Consequently, the optimal scale factors $a_i = a'_i/V_i^{1/n_i}$ follow \eqref{e:aoptgg}, and \eqref{e:dga2} is zero.
Second, we consider variations in $\bB'_i$ for any fixed (not necessarily optimal) $a'_i$, keeping $\det\! \bB'_i = 1$. Then the NSM in \eqref{e:gproda} is locally minimal when $G_i$ is locally minimal, i.e., when $\Lambda_i$ is a locally optimal lattice.
\end{IEEEproof}

Theorem \ref{t:local} and Corollary \ref{t:ropt} are curiously related to each
other: both give sufficient (but not necessary) conditions for a white
quantization error. Is Corollary \ref{t:ropt} perhaps a special case of
Theorem \ref{t:local}? In other words, is the optimal product
$\Lambda(\baopt)$, to which Theorem \ref{t:optprod} and Corollary \ref{t:ropt} apply, also a
locally optimal lattice, as defined in Sec.~\ref{s:local}?
We will see in the next section that the answer is ``no'';
for any lattice of the form \eqref{e:bproda} with $\ba=\baopt$, the NSM can be 
decreased by perturbing any of the off-diagonal submatrices written as $\bzero$ in \eqref{e:bproda}.

\begin{figure*}\centering
  \begin{subfigure}[b]{.32\textwidth}\centering

    \begin{tikzpicture}[scale=.12]
      \clip (-23,-15) rectangle (23,15);
      \fill[gray!40] (5,-6) -- (5,6) -- (-5,6) -- (-5,-6);
      \draw (0,0) node[anchor=west] {$\bzero$};
      \foreach \u in {-3,-2,-1,0,1,2,3} {\foreach \v in {-1,0,1} {
        \begin{scope}[shift={(10*\u+0*\v,12*\v)}]
          \filldraw (0,0) circle (10pt) {};
          \draw (5,-6) -- (5,6) -- (-5,6);
        \end{scope}
      }};
    \end{tikzpicture}
    \caption{$\Qopt_\Lamp$}

  \end{subfigure}
  \hfill
  \begin{subfigure}[b]{.32\textwidth}\centering

    \begin{tikzpicture}[scale=.12]
      \clip (-23,-15) rectangle (23,15);
      \fill[gray!40] (5,-6) -- (5,6) -- (-5,6) -- (-5,-6);
      \draw (0,0) node[anchor=west] {$\bzero$};
      \foreach \u in {-3,-2,-1,0,1,2,3} {\foreach \v in {-1,0,1} {
        \begin{scope}[shift={(10*\u+4*\v,12*\v)}]
          \filldraw (0,0) circle (10pt) {};
          \draw (5,-6) -- (5,6) -- (-5,6);
        \end{scope}
      }};
    \end{tikzpicture}
    \caption{$\Qsub_\Lambda$}

  \end{subfigure}
  \hfill
  \begin{subfigure}[b]{.32\textwidth}\centering

    \begin{tikzpicture}[scale=.12]
      \clip (-23,-15) rectangle (23,15);
      \fill[gray!40] (5,-5) -- (5,5) -- (-1,7) -- (-5,5) -- (-5,-5) -- (1,-7);
      \draw (0,0) node[anchor=west] {$\bzero$};
      \foreach \u in {-3,-2,-1,0,1,2,3} {\foreach \v in {-1,0,1} {
        \begin{scope}[shift={(10*\u+4*\v,12*\v)}]
          \filldraw (0,0) circle (10pt) {};
          \draw (5,-5) -- (5,5) -- (-1,7) -- (-5,5);
        \end{scope}
      }};
    \end{tikzpicture}
    \caption{$\Qopt_\Lambda$}

  \end{subfigure}
  \caption{An example of Theorem \ref{t:upper} with $n_1=n_2=1$. Each
    cell is the decision region of the lattice point it contains, and
    the shaded cells are the fundamental decision regions.  Comparing
    (a) and (b) shows that the NSMs of $\Qopt_\Lamp$ and
    $\Qsub_\Lambda$ are equal, because their fundamental decision
    regions are identical. Comparing (b) and (c) shows that
    $\Qsub_\Lambda$ cannot have a smaller NSM than $\Qopt_\Lambda$,
    because the lattices are identical and $\Qopt_\Lambda(\bx)$
    minimizes the quantization error for every $\bx$.}
  \label{f:brickwall} 
\end{figure*}
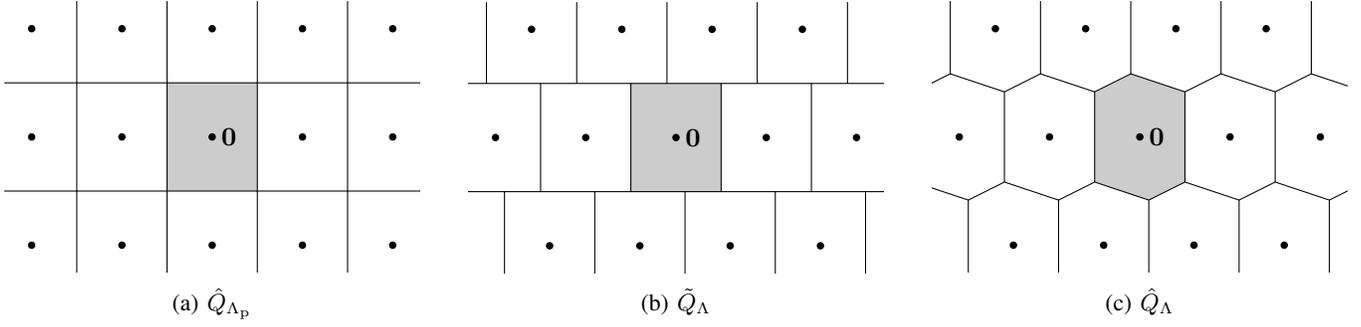

\section{Upper bound} \label{s:upper}

In this section, we show that the NSM of any lattice is bounded above by
that of a product lattice, which is given by \eqref{e:gprod} or \eqref{e:gproda}.
For brevity, we develop the case $k=2$ explicitly and then show that $k>2$ follows by induction.

Consider a square generator matrix of the form
\eq{ \label{e:bhb}
\bB = \begin{bmatrix}
  \bB_1    &    \bzero \\
  \bH    &    \bB_2
\end{bmatrix},
}
where $\bB_1$ is $n_1 \times n_1$ and $\bB_2$ is $n_2 \times n_2$.  Let
$\Lambda_1$, $\Lambda_2$, and $\Lambda$ be the lattices generated by
$\bB_1$, $\bB_2$, and $\bB$, respectively, and let $\Lamp = \Lambda_1
\times \Lambda_2$ as in Sec.~\ref{s:product}.

\begin{lemma} \label{t:upper2}
For any $\bH$, $G(\Qopt_\Lambda) \le G(\Qopt_\Lamp)$, with equality if $\bH = \bzero$.
\end{lemma}

\begin{IEEEproof}
The method of proof is to define a suboptimal quantizer
$\Qsub_\Lambda$ for which $G(\Qsub_\Lambda) = G(\Qopt_\Lamp)$. The
lemma then follows, because by the definition of $\Qopt_\Lambda$,
$G(\Qopt_\Lambda) \le G(\Qsub_\Lambda)$. The decision regions for the
three different quantizers are illustrated in Fig.~\ref{f:brickwall}.

We construct the suboptimal quantization rule
from the optimal quantization rules for $\Lambda_1$ and
$\Lambda_2$. As earlier, write the source vector as $\bx = [\bx_1 \;\;
  \bx_2]$, where $\bx_1$ and $\bx_2$ have respective dimensions $n_1$
and $n_2$. Our suboptimal quantization rule is defined by the
following four-step algorithm:
\eq{
\hat{\bx}_2 & \triangleq \Qopt_{\Lambda_2}(\bx_2) \label{e:xhat2} , \\
\bz_1 & \triangleq \hat{\bx}_2 \bB_2^{-1} \bH \label{e:z} , \\
\hat{\bx}_1 & \triangleq \Qopt_{\Lambda_1}(\bx_1 - \bz_1) , \label{e:cancellation} \\
\Qsub_\Lambda(\bx) & \triangleq [\hat{\bx}_1+\bz_1 \quad \hat{\bx}_2]. \label{e:qsub}
}
Quantities with subscript ``1'' lie in the subspace spanned by the
first $n_1$ coordinates (horizontal in Fig.~\ref{f:brickwall}) and the
quantities with subscript ``2'' lie in the subspace spanned by the
final $n_2$ coordinates (vertical in Fig.~\ref{f:brickwall}).  We
first show that this is a quantization rule: it satisfies the
conditions \eqref{e:cond1}--\eqref{e:cond2}.

Condition \eqref{e:cond1}: Since $\Qopt_{\Lambda_1}(\bx_1-\bz_1) \in \Lambda_1$
and $\Qopt_{\Lambda_2}(\bx_2) \in \Lambda_2$, there exist integer
vectors $\bu_1$ and $\bu_2$ such that $\hat{\bx}_1 = \bu_1 \bB_1$ and
$\hat{\bx}_2 = \bu_2 \bB_2$. Then by \eqref{e:z}, $\bz_1 = \bu_2 \bH$
and by \eqref{e:qsub}, $\Qsub_\Lambda(\bx) = [\bu_1\bB_1+\bu_2\bH \;\;
  \bu_2\bB_2] = [\bu_1 \;\; \bu_2]\bB$, which shows that
$\Qsub_\Lambda(\bx) \in\Lambda$.

Condition \eqref{e:cond2}: Consider $\Qsub_\Lambda(\bx+\blambda)$ for an
arbitrary $\blambda \in \Lambda$. Let $\blambda \triangleq [\bv_1 \;\;
  \bv_2] \bB = [\bv_1 \bB_1+\bv_2 \bH \;\; \bv_2 \bB_2]$, where
$\bv_1$ and $\bv_2$ are integer vectors. Let $\bx' \triangleq [\bx'_1
  \;\; \bx'_2] \triangleq \bx+\blambda$, and let $\hat{\bx}'_2$,
$\bz_1'$, $\hat{\bx}'_1$, and $\Qsub_\Lambda(\bx')$ denote the
quantities obtained via the algorithm \eqref{e:xhat2}--\eqref{e:qsub}
when $\bx$ is replaced by $\bx'$.  Since
$\Qopt_{\Lambda_1}$ and $\Qopt_{\Lambda_2}$ satisfy \eqref{e:cond2}, one finds
\eq{
\hat{\bx}'_2
  &= \Qopt_{\Lambda_2}(\bx_2+\bv_2 \bB_2) \notag\\
  &= \hat{\bx}_2+\bv_2 \bB_2 , \\
\bz_1'
  &= (\hat{\bx}_2+\bv_2\bB_2) \bB_2^{-1} \bH \notag\\
  &= \bz_1+\bv_2 \bH , \\
\hat{\bx}'_1
  &= \Qopt_{\Lambda_1}(\bx_1+\bv_1 \bB_1+\bv_2 \bH - \bz_1-\bv_2 \bH) \notag\\
  &= \hat{\bx}_1+\bv_1 \bB_1 , \\
\Qsub_\Lambda(\bx')
  &= [\hat{\bx}_1+\bv_1 \bB_1+\bz_1+\bv_2 \bH \quad \hat{\bx}_2+\bv_2 \bB_2] \notag\\
  &= \Qsub_\Lambda(\bx) + \blambda .
}
Thus, since it satisfies both conditions, $\Qsub_\Lambda$ is a valid quantization rule.

The NSM of $\Qsub_\Lambda$ is determined by its fundamental decision
region, defined by \eqref{e:vqdef}. From \eqref{e:qsub} and \eqref{e:z},
$\Qsub_\Lambda(\bx) = \bzero$ if and only if $\hat{\bx}_1 = \bzero$
and $\hat{\bx}_2 = \bzero$. Hence, \eqref{e:vqdef} yields
\eq{
 \Omega(\Qsub_\Lambda) &= \{ \bx \in \R^n: \Qopt_{\Lambda_1}(\bx_1) = \bzero \text{ and } \Qopt_{\Lambda_2}(\bx_2) = \bzero \} \notag\\
  &=  \Omega(\Qopt_{\Lambda_1}) \times  \Omega(\Qopt_{\Lambda_2}) \notag\\
  &=  \Omega(\Qopt_\Lamp) .
}
Thus, the fundamental decision region of the (suboptimal) quantization
rule for $\Lambda$ is identical to the fundamental decision region of
the optimal quantization rule for $\Lamp$. This can be intuitively
understood by comparing Figs.~\ref{f:brickwall}(a) and
\ref{f:brickwall}(b).

Since the fundamental decision regions of $\Qsub_\Lambda$ and
$\Qopt_\Lamp$ are identical, so are all properties derived from these
regions, e.g., $E(\Qsub_\Lambda) = E(\Qopt_\Lamp)$,
$\bR(\Qsub_\Lambda) = \bR(\Qopt_\Lamp)$, and $G(\Qsub_\Lambda) =
G(\Qopt_\Lamp)$.  But since the optimal decision rule for $\Lambda$
satisfies $G(\Qopt_\Lambda) \le G(\Qsub_\Lambda)$, our proof is
complete: $G(\Qopt_\Lambda) \le G(\Qopt_\Lamp)$. Equality if and only
if $\bH = \bzero$ follows, since $\bB$ in \eqref{e:bhb} and $\bBp$ in
\eqref{e:bp} are equal if and only if $\bH = \bzero$.
\end{IEEEproof}

In digital communications, the suboptimal quantization rule $\Qsub_\Lambda$ is
known as \emph{successive interference cancellation} \cite{wubben11}.
In that scenario, $\bx_1$ and $\bx_2$ represent information received on two
parallel communication channels, which interfere with each other.
If $\hat{\bx}_2$ is detected first \eqref{e:xhat2}, its effect on $\bx_1$ can be calculated
\eqref{e:z} and cancelled \eqref{e:cancellation} before
$\hat{\bx}_1$ is detected in \eqref{e:qsub}.
If \eqref{e:xhat2}--\eqref{e:qsub} are extended to $n$ steps of
one-dimensional quantization, then the resulting suboptimal quantization rule
yields the \emph{Babai point} \cite{babai86}.

The extension to $k>2$ follows immediately.
Let $\Lambda$ be the lattice generated by
\eq{ \label{e:bhbhb}
\bB = \begin{bmatrix}
  \bB_1 & \bzero & \cdots & \bzero \\
\bH_{2,1} & \bB_2 & \cdots & \bzero \\
\vdots & \vdots & \ddots & \vdots \\
\bH_{k,1} & \bH_{k,2} & \cdots & \bB_k
\end{bmatrix} ,
}
$\Lambda_i$ be the lattice generated by $\bB_i$ for $i=1,\ldots,k$,
and $\Lamp = \Lambda_1 \times \cdots \times \Lambda_k$.
\begin{theorem} \label{t:upper}
For given $\bB_1,\ldots,\bB_k$ and any $\bH_{i,j}$,
$G(\Qopt_\Lambda) \le G(\Qopt_\Lamp)$, with equality if $\bH_{i,j} = \bzero$ for all $i,j$.
\end{theorem}

\begin{IEEEproof}
By induction. If the theorem holds for the upper-left $(k-1)\times(k-1)$ part of \eqref{e:bhbhb},
then it extends to $k\times k$ by Lemma \ref{t:upper2}.
\end{IEEEproof}

Like Proposition \ref{t:prod}, Theorem \ref{t:upper} can also be extended
by scale factors $\ba$. Specifically, if $\bB_1, \ldots, \bB_k$ in \eqref{e:bhbhb} are
multiplied by scale factors $\ba = [a_1, \ldots, a_k]$, then the NSM $G(\ba)$ of the resulting
lattice is bounded by \eqref{e:gproda} for arbitrary given $\ba$
or by \eqref{e:goptgg} for optimal $\ba$.

We now return to the question of whether the optimal product $\Lambda(\ba)$
generated by $\bB(\ba)$ in \eqref{e:bproda} is always locally optimal at $\ba=\baopt$.
It was observed in Corollary~\ref{t:ropt} that if $\Lambda_1,\ldots,\Lambda_k$ are locally optimal,
then $G(\baopt)$ is locally
\emph{minimal} with respect to perturbations in any submatrix $a_i \bB_i$ about its local optimum.
On the other hand, it follows from Theorem \ref{t:upper}
that $G(\baopt)$ is locally \emph{maximal} with respect to perturbations
about $\bzero$ in any submatrix below the block diagonal of $\bB(\baopt)$. 
The same theorem holds if \eqref{e:bhbhb} is replaced by an upper-triangular 
matrix $\bB$.
This proves that $G(\baopt)$ is also locally maximal with
respect to perturbations about $\bzero$ \emph{above} the block diagonal of $\bB(\baopt)$.
Since the NSM increases for any perturbations in the submatrices on the diagonal
of \eqref{e:bproda} and decreases for any perturbations 
in submatrices either below or above the block diagonal, 
the first derivative of $G$ with
respect to these entries must vanish:
the NSM has a \emph{saddle
point} at $\bB(\baopt)$. In conclusion, all lattices that fulfill \eqref{e:white}
are \emph{not} locally optimal.

One way to construct lattices is by \emph{lamination} of a lower-dimensional
lattice.  To build an $n$-dimensional laminated lattice $\Lambda$,
take a generator matrix $\bB_1$ for an $(n-1)$-dimensional lattice
$\Lambda_1$ and an arbitrary $(n-1)$-dimensional vector $\bh$. Then
construct the $n\times n$ generator matrix
\begin{equation}
  \label{e:lamination}
\bB = \begin{bmatrix}
  \bB_1    &    \bzero \\
  \bh    &    a
\end{bmatrix}.
\end{equation}
Here $a > 0 $ is a real number, which is the distance between the
shifted lattice copies in the direction orthogonal to their subspace,
and the vector $\bh$ is the stacking offset in the $(n-1)$-plane.
Fig.~\ref{f:topology} illustrates the Voronoi region of a laminated
lattice with $n=3$ and $\bh = \bzero$.

In the classical lattice literature, ``laminated lattices'' are built
recursively in this way, to maximize the packing density, starting
from $\Z$. So $\bB_1$ is itself the generator for a laminated lattice,
and in each recursive iteration, $\bh$ and $a$ in \eqref{e:lamination}
are selected to maximize the packing density.  This construction gives
rise to the well-studied $\Lambda$ and $K$ series in
\cite{conway82laminated, plesken93}, \cite[Sec.~4 of Ch.~5 and
  Ch.~6]{splag}.

In this paper, our focus is the NSM rather than the packing density,
so we use ``laminated lattice'' more broadly for \emph{any} lattice generated
via \eqref{e:lamination}. This broader meaning is consistent with
\cite{dutour08, allen21}.  Note that to maximize the packing density,
the optimal choice of $\bh$ is a ``deep hole'' in $\Lambda_1$ (a
vertex of the Voronoi region most distant from the origin). This is
also (intuitively) a good choice to minimize the NSM, although it may
not be optimal.

An upper bound on the quantization performance of a laminated lattice
follows directly from the results of Secs.~\ref{s:product} and
\ref{s:upper}, as follows.
\begin{corollary} \label{t:lam}
An $n$-dimensional lattice $\Lambda$ obtained by lamination of an
$(n-1)$-dimensional lattice $\Lambda_1$ satisfies, for an arbitrary
offset $\bh$ and the optimal layer separation $a$,
\eq{G(\Qopt_\Lambda) \le
  \frac{G(\Qopt_{\Lambda_1})^{1-\frac{1}{n}}}{12^{\frac{1}{n}}} \label{e:lam}
  .  }
\end{corollary}

\begin{IEEEproof}
Setting $n_2=1$, $\Lambda_2 = a\Z$, and $\bH=\bh$ in Lemma
\ref{t:upper2} yields
\eq{
G(\Qopt_\Lambda) \le G(\Qopt_\Lamp) \label{e:glam1} ,
}
where $\Lamp = \Lambda_1 \times a\Z$. Furthermore, setting $n_2=1$,
$\Lambda_2 = a\Z$, $G_1 = G(\Qopt_{\Lambda_1})$, and $G_2 =
G(\Qopt_\Z) = 1/12$ in Theorem \ref{t:optprod} yields
\eq{
G(\Qopt_\Lamp) \le \frac{G(\Qopt_{\Lambda_1})^{1-\frac{1}{n}}}{12^{\frac{1}{n}}} \label{e:glam2} .
}
Combining \eqref{e:glam1} and \eqref{e:glam2} completes the proof.
\end{IEEEproof}

As a curiosity, we note that \eqref{e:lam} would have been more appealing if $G$ had been defined a factor of $12$ larger than
the standard definition \eqref{e:gqdef}. With that alternative definition,
$\Z^n$ would have an NSM
of $1$ in any dimension and the denominator would disappear from expressions like \eqref{e:lam} and \eqref{e:glam2}.

\begin{table*}
\begin{center}
\caption{The best known lattice quantizers. Columns 2--3 list the
  smallest normalized second moments (NSM) $G$ previously reported in
  dimension $n$, while columns 4--5 list the conjectured lower and
  upper bounds, respectively. Columns 6--7 list our best product
  lattices.  Columns 8--9 indicate if the product lattice in Column 7
  provides a smaller NSM than previously reported in Column 2 ($<$G) and/or is
  below the upper bound of Column 5 ($<$U).  NSMs that are are known
  exactly are listed with nine decimals, whereas NSMs with five
  decimals are derived from numerical estimates in \cite{conway84}.}
\label{t:bestquantizers}
\begin{tabular}{r|ll|ll|llll}
\hline
\tstrut & \multicolumn{2}{c|}{Best previously reported} & \multicolumn{2}{c|}{Generic bounds} & \multicolumn{4}{c}{Best product} \\
\bstrut $n$ & \;\;\;\;\;\;NSM & \!\!\!\!Lattice & \;\;Lower \cite{conway85} & \;\;Upper \cite{torquato10} & \;\;\;\;\;\;NSM & \;\;\;\;Lattice &  \multicolumn{2}{c}{Better?}  \\
\hline\hline
\tstrut$1$ & $0.083333333$ & $\Z$ & $0.083333333$ & $0.083333333$ & & & &\\
$2$ & $0.080187537$ & $A_{2}$ & $0.080187537$ & $0.080267223$ & $0.083333333$ & $\Z \otimes \Z$ & &\\
$3$ & $0.078543281$ & $A^*_{3}$ & $0.077874985$ & $0.079723711$ & $0.081222715$ & $A_{2} \otimes \Z$ & &\\
$4$ & $0.076603235$ & $D_{4}$ & $0.076087080$ & $0.078822518$ & $0.079714343$ & $A^*_{3} \otimes \Z$ & &\\
$5$ & $0.075625443$ & $D^*_{5}$ & $0.074654327$ & $0.078731261$ & $0.077904301$\footnotemark & $D_{4} \otimes \Z$ & & $<$U\\
$6$ & $0.074243697$ & $E^*_{6}$ & $0.073474906$ & $0.077779975$ & $0.076858706$ & $D^*_{5} \otimes \Z$ & & $<$U\\
$7$ & $0.073116493$ & $E^*_{7}$ & $0.072483503$ & $0.076858058$ & $0.075478834$ & $E^*_{6} \otimes \Z$ & & $<$U\\
$8$ & $0.071682099$ & $E_{8}$ & $0.071636064$ & $0.075654034$ & $0.074321725$ & $E^*_{7} \otimes \Z$ & & $<$U\\
\hline
$9$ & $0.071622594$ & $\mathit{AE}_{9}$ & $0.070901661$ & $0.075552237$ & $0.072891732$ & $E_{8} \otimes \Z$ & & $<$U\\
$10$ & $0.070813818$ & $D^+_{10}$ & $0.070257874$ & $0.074856027$ & $0.072715487$ & $\mathit{AE}_{9} \otimes \Z$ & & $<$U\\
$11$ & $0.070426259$ & $A_{11}^{3}$ & $0.069688002$ & $0.074026177$ & $0.071869620$ & $D^+_{10} \otimes \Z$ & & $<$U\\
$12$ & $0.070095600$ & $K_{12}$ & $0.069179323$ & $0.073098569$ & $0.071420842$ & $A_{11}^{3} \otimes \Z$ & & $<$U\\
$13$ & $0.074873919$ & $D^*_{13}$ & $0.068721956$ & $0.072400247$ & $0.071034583$ & $K_{12} \otimes \Z$ & $<$G & $<$U\\
$14$ & $0.074954492$ & $D^*_{14}$ & $0.068308096$ & $0.071672217$ & $0.071455542$ & $K_{12} \otimes A_{2}$ & $<$G & $<$U\\
$15$ & $0.075039738$ & $D^*_{15}$ & $0.067931488$ & $0.071008692$ & $0.071709124$ & $K_{12} \otimes A^*_{3}$ & $<$G &\\
$16$ & $0.06830$ & $\Lambda_{16}$ & $0.067587055$ & $0.070399705$ & $0.071668753$ & $K_{12} \otimes D_{4}$ & &\\
\hline
$17$ & $0.075216213$ & $D^*_{17}$ & $0.067270625$ & $0.069886791$ & $0.06910$ & $\Lambda_{16} \otimes \Z$ & $<$G & $<$U\\
$18$ & $0.075304924$ & $D^*_{18}$ & $0.066978741$ & $0.069403282$ & $0.06953$ & $\Lambda_{16} \otimes A_{2}$ & $<$G &\\
$19$ & $0.075392902$ & $D^*_{19}$ & $0.066708503$ & $0.068958664$ & $0.06982$ & $\Lambda_{16} \otimes A^*_{3}$ & $<$G &\\
$20$ & $0.075479665$ & $D^*_{20}$ & $0.066457468$ & $0.068548490$ & $0.06988$ & $\Lambda_{16} \otimes D_{4}$ & $<$G &\\
$21$ & $0.075554858$ & $A^*_{21}$ & $0.066223553$ & $0.068173104$ & $0.06998$ & $\Lambda_{16} \otimes D^*_{5}$ & $<$G &\\
$22$ & $0.075577414$ & $A^*_{22}$ & $0.066004976$ & $0.067826205$ & $0.06987$ & $\Lambda_{16} \otimes E^*_{6}$ & $<$G &\\
$23$ & $0.075601888$ & $A^*_{23}$ & $0.065800200$ & $0.067506055$ & $0.06973$ & $\Lambda_{16} \otimes E^*_{7}$ & $<$G &\\
$24$ & $0.06577$ & $\Lambda_{24}$ & $0.065607893$ & $0.067208741$ & $0.06941$ & $\Lambda_{16} \otimes E_{8}$ & &\\
\hline
$25$ & $0.075655156$ & $A^*_{25}$ & $0.065426891$ & $0.066939780$ & $0.06640$ & $\Lambda_{24} \otimes \Z$ & $<$G & $<$U\\
$26$ & $0.075683386$ & $A^*_{26}$ & $0.065256179$ & $0.066685133$ & $0.06678$ & $\Lambda_{24} \otimes A_{2}$ & $<$G &\\
$27$ & $0.075712385$ & $A^*_{27}$ & $0.065094858$ & $0.066446178$ & $0.06708$ & $\Lambda_{24} \otimes A^*_{3}$ & $<$G &\\
$28$ & $0.075741975$ & $A^*_{28}$ & $0.064942137$ & $0.066222221$ & $0.06722$ & $\Lambda_{24} \otimes D_{4}$ & $<$G &\\
$29$ & $0.075772009$ & $A^*_{29}$ & $0.064797312$ & $0.066012092$ & $0.06737$ & $\Lambda_{24} \otimes D^*_{5}$ & $<$G &\\
$30$ & $0.075802366$ & $A^*_{30}$ & $0.064659756$ & $0.065814436$ & $0.06738$ & $\Lambda_{24} \otimes E^*_{6}$ & $<$G &\\
$31$ & $0.075832940$ & $A^*_{31}$ & $0.064528911$ & $0.065628191$ & $0.06736$ & $\Lambda_{24} \otimes E^*_{7}$ & $<$G &\\
$32$ & $0.075863646$ & $A^*_{32}$ & $0.064404271$ & $0.065452387$ & $0.06720$ & $\Lambda_{24} \otimes E_{8}$ & $<$G &\\
\hline
$33$ & $0.075894409$ & $A^*_{33}$ & $0.064285386$ & $0.065286162$ & $0.06732$ & $\Lambda_{24} \otimes \mathit{AE}_{9}$ & $<$G &\\
$34$ & $0.075925169$ & $A^*_{34}$ & $0.064171846$ & $0.065128681$ & $0.06722$ & $\Lambda_{24} \otimes D^+_{10}$ & $<$G &\\
$35$ & $0.075955874$ & $A^*_{35}$ & $0.064063282$ & $0.064979253$ & $0.06720$ & $\Lambda_{24} \otimes A_{11}^{3}$ & $<$G &\\
$36$ & $0.075986480$ & $A^*_{36}$ & $0.063959359$ & $0.064837252$ & $0.06718$ & $\Lambda_{24} \otimes K_{12}$ & $<$G &\\
$37$ & $0.076016949$ & $A^*_{37}$ & $0.063859771$ & $0.064702115$ & $0.06757$ & $\Lambda_{24} \otimes K_{12} \otimes \Z$ & $<$G &\\
$38$ & $0.076047252$ & $A^*_{38}$ & $0.063764240$ & $0.064573335$ & $0.06781$ & $\Lambda_{24} \otimes K_{12} \otimes A_{2}$ & $<$G &\\
$39$ & $0.076077363$ & $A^*_{39}$ & $0.063672511$ & $0.064450455$ & $0.06799$ & $\Lambda_{24} \otimes K_{12} \otimes A^*_{3}$ & $<$G &\\
$40$ & $0.076107259$ & $A^*_{40}$ & $0.063584352$ & $0.064333062$ & $0.06677$ & $\Lambda_{24} \otimes \Lambda_{16}$ & $<$G &\\
\hline
$41$ & $0.076136923$ & $A^*_{41}$ & $0.063499548$ & $0.064220781$ & $0.06713$ & $\Lambda_{24} \otimes \Lambda_{16} \otimes \Z$ & $<$G &\\
$42$ & $0.076166341$ & $A^*_{42}$ & $0.063417902$ & $0.064113272$ & $0.06736$ & $\Lambda_{24} \otimes \Lambda_{16} \otimes A_{2}$ & $<$G &\\
$43$ & $0.076195500$ & $A^*_{43}$ & $0.063339234$ & $0.064010223$ & $0.06753$ & $\Lambda_{24} \otimes \Lambda_{16} \otimes A^*_{3}$ & $<$G &\\
$44$ & $0.076224390$ & $A^*_{44}$ & $0.063263376$ & $0.063911352$ & $0.06761$ & $\Lambda_{24} \otimes \Lambda_{16} \otimes D_{4}$ & $<$G &\\
$45$ & $0.076253004$ & $A^*_{45}$ & $0.063190174$ & $0.063816399$ & $0.06770$ & $\Lambda_{24} \otimes \Lambda_{16} \otimes D^*_{5}$ & $<$G &\\
$46$ & $0.076281336$ & $A^*_{46}$ & $0.063119483$ & $0.063725126$ & $0.06770$ & $\Lambda_{24} \otimes \Lambda_{16} \otimes E^*_{6}$ & $<$G &\\
$47$ & $0.076309381$ & $A^*_{47}$ & $0.063051171$ & $0.063637315$ & $0.06768$ & $\Lambda_{24} \otimes \Lambda_{16} \otimes E^*_{7}$ & $<$G &\\
\bstrut$48$ & $0.076337136$ & $A^*_{48}$ & $0.062985115$ & $0.063552764$ & $0.06577$ & $\Lambda_{24} \otimes \Lambda_{24}$ & $<$G &\\
\hline
\multicolumn{7}{l}{\!\!\!\!\parbox{8cm}{
\overlay{-1.7cm}{-1.4cm}{  %
    \footnotemark[\value{footnote}]Proposed as an upper bound in \cite[Tab.~I]{gersho79}.
}
}}
\end{tabular}
\end{center}
\end{table*}

\section{Best Known Lattice Quantizers}

The upper bound in Corollary \ref{t:lam} has interesting
implications. These call to mind an observation made by Cohn in the
context of sphere packing \cite{cohn17}.  Referring to a plot of
sphere-packing density as a function of dimension, he comments that
``\emph{Certain dimensions, most notably 24, have packings so good
that they seem to pull the entire curve in their direction. The fact
that this occurs is not so surprising, since one expects cross
sections and stackings of great packings to be at least good, but the
effect is surprisingly large.}''
Indeed, the sphere-packing density in a given dimension $n$ is bounded by a function
of the sphere-packing density in dimension $n-1$ according to
\emph{Mordell's inequality} \cite[Eq.~(19) of Ch.~6]{splag}.
Here, in the context of lattice
quantizers, Corollary \ref{t:lam} does precisely this for NSMs.  A lattice with
particularly small NSM $G$ in dimension $n-1$ makes it possible to
also obtain a small NSM in dimension $n$, and hence ``pulls down the
NSM curve'' for larger dimensions. More generally, Theorem \ref{t:upper}
can pull down the curve over intervals of more than one dimension.

We designed product lattices in dimension $n$ by applying Theorem
\ref{t:optprod} with $k=2$ to the best known lattices in dimensions $n_1$ and
$n_2=n-n_1$, for $n_1$ ranging from $1$ to
$n-1$. 
There is no need to explicitly consider $k>2$, although
recursive application of Theorem \ref{t:optprod} with $k=2$
can generate product lattices with larger $k$.
The minimal NSM obtained in each dimension provides a
constructive upper bound on the optimal NSM. It follows from Theorem
\ref{t:upper} that better lattice quantizers can be found among
lattices of the form \eqref{e:bhbhb}, where $\bB_1,\ldots,\bB_k$ are the
best known lattices in their dimensions, for
example by lamination if $n_i = 1$ for any $i$.

Tab.~\ref{t:bestquantizers} summarizes the best known lattice
quantizers in dimensions $n \le 48$. The first such list was compiled
in 1979 for $n=1$ to $5$ \cite{gersho79}. It was extended to $n\le 10$
in 1982 \cite{conway82voronoi}, which also analytically calculated the NSMs of the classical
lattices $A_n$, $A^*_n$, $D_n$, and $D^*_n$
for any $n$.  Since then, progress has been much slower. Better
lattices were reported for $n=6$ and $7$ in \cite{conway84} and for
$n=9$ and $10$ in \cite{agrell98}, although the NSMs of these lattices
were only computed numerically. Reference \cite{conway84} also gave
the best known lattice quantizers for $n=12$, $16$, and $24$ with
numerically computed NSMs. Later, the corresponding exact NSMs were
calculated for $n=6$ \cite{worley87}, $n=7$ \cite{worley88}, $n=9$
\cite{allen21}, and $n=10$ and $12$ \cite{dutour09}.  In the latter
reference, a new best known lattice quantizer was identified for
$n=11$. NSM results for $n \le 15$ were summarized in
\cite{PhysRevD.104.042005}.

In Tab.~\ref{t:bestquantizers}, we also include the best known lower and upper
bounds on the NSM.
The lower bound is a conjecture by Conway and Sloane \cite{conway85}, which we evaluated numerically by high-resolution trapezoidal integration.
The upper bound is by Torquato in
\cite[Eq.~(105)]{torquato10}\footnote{See \cite[Note 31]{allen22} for corrections.}
and improves on a well-known bound by Zador \cite[Lemma 5]{zador82}.
The Torquato upper bound is for arbitrary
quantizers, which might not be lattices.
We conjecture that there is always at least one lattice
quantizer satisfying this bound. For the best known packing densities, which
are needed to evaluate \cite[Eq.~(105)]{torquato10}, we use \cite[Tab.~I.1]{splag}.
Taken together, these two results provide
conjectured lower and upper bounds on the NSM of optimal lattice
quantizers.  While intuitively plausible, these remain unproven.

For each dimension, the last four columns of
Tab.~\ref{t:bestquantizers} list the best product lattice that can be
constructed from the lattices given on the preceding rows of the
table. While Theorem \ref{t:upper} establishes that these lattices are
not even locally optimal, they nevertheless improve significantly on
the previously lowest reported NSMs. For brevity, we use $\otimes$ to
denote the Cartesian product of lattices {\it with the optimal
  choice of relative scale}. More precisely, $\Lambda_1 \otimes
\cdots\otimes\Lambda_k \triangleq a_1\Lambda_1 \times\cdots\times a_k\Lambda_k$ with
$[a_1,\ldots,a_k]$
given by Theorem \ref{t:optprod}.

The first dimension in which this construction provides a better
lattice quantizer than previously reported is $n=13$, where
our best product lattice is $K_{12}\otimes \Z$. With an NSM of
$0.0710$, it is the first reported $13$-dimensional lattice whose NSM
falls below the generic upper bound $0.0724$. It is significantly
better than the currently best known $13$-dimensional lattice
quantizer $D^*_{13}$, whose NSM is $0.0749$.  Our optimized product
lattices are also the first reported lattices which lie below the
upper bound in dimensions $n=14$, $17$, and $25$.

\section{Conclusions}

As far as we know, Table~\ref{t:bestquantizers} is currently the most
extensive published table of record NSM lattices.%
\footnote{Several improved lattice quantizers in dimensions $n\le 24$ are
reported in \cite{lyu22}, which appeared while this
paper was in review.}
Previous tables have
not gone beyond $n=16$, apart from a numerical estimate for $n=24$.
It is disconcerting that in many dimensions the smallest known NSMs
are for product lattices, since we have proven that these cannot
be optimal.  This reflects the complexity of designing good lattices
and of evaluating their NSMs. The theoretical results in this paper
provide some properties of optimal lattice quantizers, which in
combination with the tabulated product lattices may hopefully provide
guidance and benchmarks for further progress in lattice quantization.
An analytic or algorithmic method to optimize $\bH$ in \eqref{e:bhb}
or $\bh$ in \eqref{e:lamination} would be highly desirable.

\balance

\end{document}